\documentclass[twocolumn,prd,letterpaper,amsmath,amssymb,nofootinbib]{revtex4}
\usepackage{amsmath,amssymb,graphicx}
\usepackage {amssymb}
\usepackage{multirow}
\usepackage{verbatim}

\newcommand{\nc}{\newcommand}
\nc{\ba}{\begin{eqnarray}}
\nc{\ea}{\end{eqnarray}}
\nc{\be}{\begin{eqnarray}}
\nc{\ee}{\end{eqnarray}}

\newcommand\s{\sigma}

\newcommand\ta{\tau}

\nc{\ga}{\gamma}
\nc{\om}{\omega}

\nc{\x}{{\bf x }}
\nc{\kk}{{\bf k }}
\nc{\f}{{\bf f }}
\nc{\e}{{\bf e }}
\nc{\T}{ \theta (s_i (t)- \s) }
\nc{\TT}{ \theta (s_i (t_{ r \, i } )- \s) }
\nc{\br}{   (s_i (t)- \s)  }
\input{epsf.sty}

\newcommand{\til}{\tilde}

\newcommand{\gsim}{ \lower .75ex \hbox{$\sim$} \llap{\raise .27ex \hbox{$>$}} }
\newcommand{\lsim}{ \lower .75ex \hbox{$\sim$} \llap{\raise .27ex \hbox{$<$}} }

\begin{document}

\title{Preheating with the Brakes On: The Effects of a Speed Limit}

\author{Johanna Karouby, Bret Underwood, Aaron C.~Vincent}
\affiliation{Department of Physics, McGill University\\
3600 University Street, Montr\'eal, Qu\'ebec, Canada H3A 2T8
}

\begin{abstract}
We study preheating in models where the inflaton has a non-canonical kinetic term, containing powers of the usual 
kinetic energy.  The inflaton field oscillating about its potential minimum acts as a driving force for 
particle production through parametric resonance.
Non-canonical kinetic terms can impose a speed limit on the motion of the inflaton, modifying 
the oscillating inflaton profile.  This has two important effects: it turns a smooth sinusoidal profile
into a sharp saw-tooth, enhancing resonance, and it lengthens the period of oscillations, suppressing resonance.
We show that the second effect dominates over the first, so that preheating with a non-canonical inflaton field
is less efficient than with canonical kinetic terms, and that the expansion of the Universe suppresses resonance even further.

\end{abstract}

\maketitle

\section{Introduction}

After a sufficiently long period of inflation the Universe would be cold and devoid of observable matter.
The energy responsible for driving inflation is trapped in the (nearly) homogeneous inflaton field $\phi$.
In order for observable matter to emerge from the post-inflationary Universe, the inflaton field must couple
to additional degrees of freedom in a way that the inflationary energy is dumped into observable matter
through a process known as reheating
\cite{Dolgov:1982th,Abbott:1982hn,preheatingLong}.  
For the purpose of reheating, the inflaton can couple to other scalar fields, fermions, or gauge fields.

If the inflaton couples to bosonic fields, such as other scalar fields, novel
condensation effects can take place.  In particular, because there is no exclusion principle, the inflaton field can
transfer a large amount of energy to 
the reheating field $\chi$ (the ``reheaton")
in a process that is far from
equilibrium.  Such enhancements result from non-linear resonance effects due to the interaction between the
inflaton and reheaton.  For example, for the Lagrangian
\ba
{\mathcal L} = \frac{1}{2}(\partial \phi)^2 - V(\phi) + \frac{1}{2}(\partial \chi)^2 + \frac{1}{2}g^2 \phi^2 \chi^2\, ,
\ea
the equation of motion for large-scale modes of $\chi$ (neglecting the expansion of the Universe) becomes that of a
harmonic oscillator with a time-dependent frequency,
\ba
\ddot \chi(t) + \left(k^2 + g^2 \phi(t)^2\right)\chi(t) = 0
\label{eq:chieomintro}
\ea
where $k$ is the comoving wavenumber of the $\chi$ field and $\phi = \phi(t)$ is the time-dependent background
solution for the inflaton.  At the end of single field inflation the inflaton oscillates about the minimum of its potential,
so that the time-dependent frequency $\omega(t)^2 = k^2 + g^2 \phi(t)^2$ oscillates with time.
It is well-known that a harmonic oscillator with an oscillating time-dependent frequency can exhibit resonance effects,
where the amplitude $\chi(t)$ of the oscillator grows exponentially with time.  This post-inflationary 
exponential growth of
certain long-wavelength modes of $\chi$ is dubbed ``preheating" 
\cite{Dolgov:1989us,Traschen:1990sw,preheating,Shtanov:1994ce,preheatingLong,PreheatExpanding}, 
and has been studied in a variety
of different contexts. See 
\cite{InflationReview,Kofman:2008zz,ReheatReview} 
for some reviews of this extensive literature.
Most importantly for this paper, the majority of these studies (with the exception of \cite{DBIPreheat2})
have focused only on quadratic kinetic terms for the effective theory of reheating.

This approach towards inflationary and reheating model building neglects, however, the fact that these simple 
Lagrangians are really effective field theory (EFT) descriptions, only valid at sufficiently low energies.  
In particular, these Lagrangians
should be understood as having been obtained by integrating out physics above some scale $\Lambda$
at which new physics (such as new fields or new interactions) become important.
The effects of physics above this energy can be parameterized in the EFT through non-renormalizable operators
suppressed by powers of the scale of new physics,
\ba
{\mathcal L}_{eff} = {\mathcal L}_0 + \sum_{n>4} c_n \frac{{\mathcal O}_n}{\Lambda^{n-4}}\, .
\label{eq:effL}
\ea
As an example, consider the two-field Lagrangian,
\ba
{\mathcal L}= \frac{1}{2}(\partial \phi)^2 + \frac{1}{2}(\partial \rho)^2 + \frac{\rho}{\Lambda}(\partial \phi)^2 - \frac{1}{2} \Lambda^2 \rho^2 - V(\phi)\, ,
\ea
with $\phi$ the inflaton field and $\rho$ some heavy field with mass $\Lambda$.  For energies below $\Lambda$, we
can integrate out $\rho$ at the classical (tree) level to obtain the effective Lagrangian 
\cite{Franche:2009gk,Gelaton},
\ba
{\mathcal L}_{eff} = \frac{1}{2} (\partial \phi)^2 + \frac{(\partial \phi)^4}{\Lambda^4} - V(\phi)\,.
\ea
The effective theory now contains a new contribution to the kinetic part of the action for the inflaton.
More generally, one can consider as a low energy EFT a Lagrangian for the inflaton of the form\footnote{This is, 
of course, not the most general effective theory of the form (\ref{eq:effL}).  In particular, we have omitted terms
involving higher derivatives.  This can be done self-consistently as long as the higher derivatives are small
for the physics we are interested in, which we will argue is the case.  See also \cite{Franche:2009gk} 
for further analysis of the validity of the truncation
(\ref{eq:LeffNonCanon}) in the context of inflation.  It is also worthwhile to note that certain Lagrangians of the
form (\ref{eq:LeffNonCanon}) are protected against corrections by powerful non-linear symmetries 
\cite{deRham:2010eu}.}
\ba
{\mathcal L}_{eff} = {\mathcal L}_{eff}(X,\phi)\, ,
\label{eq:LeffNonCanon}
\ea
where $X \equiv -\frac{1}{2} (\partial \phi)^2$.  Inflation with Lagrangians of this type can have novel
features, such as a speed limit on the motion of the homogeneous inflaton and 
a sound speed of perturbations less than one $c_s^2 \leq 1$, 
that have important implications for
inflationary perturbations and models \cite{kinflation,kinflationPert,DBI,DBISky,NonGauss}.

It is the post-inflationary dynamics of (\ref{eq:LeffNonCanon}) that is most interesting to us here.
In particular, we will couple the non-canonical inflaton in (\ref{eq:LeffNonCanon}) to a canonical reheaton field through
a quartic interaction:
\be
{\mathcal L}_{pre} = {\mathcal L}_{eff}(X,\phi) + \frac{1}{2} (\partial \chi)^2 
  - \frac{1}{2} g^2 \phi^2 \chi^2 - \frac{1}{2} m_\chi^2 \chi^2\, .
\label{eq:twosector}
\ee
The equation of motion for the reheaton is still of the form (\ref{eq:chieomintro}).  However, since
the resonance arising from the time-dependent harmonic oscillator (\ref{eq:chieomintro}) is a non-linear effect,
it is sensitive to the precise profile of the inflaton $\phi(t)$ as it oscillates about its potential minimum. 
The profile of the oscillation is in turn sensitive to the non-canonical kinetic terms in (\ref{eq:twosector}).
Thus, the modified dynamics from non-canonical Lagrangians can play an important role in the physics of preheating.

In this paper, we will discuss the implications of an inflaton sector with non-canonical kinetic terms for preheating.
We first discuss in Section \ref{sec:noncanonKin} the dynamics of a non-canonical inflaton oscillating about
its minimum.  A fairly generic feature is the existence of a speed limit on the motion of the inflaton, restricting how
fast it can move in field space.  When the oscillating inflaton saturates this speed limit, its profile is no longer sinusoidal
but instead takes a saw-tooth form.
In Section \ref{sec:ncpreheating} we show how the equation of motion for reheaton perturbations can be
recast into a form of Hill's equation, for which well-known techniques exist for finding resonance bands.
Two competing effects lead to modifications of the standard theory of preheating: non-canonical kinetic terms
lead to a ``sharper" inflaton profile, enhancing resonance, while the period of oscillation is lengthened due to the
speed limit, suppressing resonance.  The net result is that the latter effect dominates, so that the resonance
for non-canonical kinetic terms is {\it less efficient} than its canonical counterpart.
After illustrating the effects of the expansion of the Universe, we summarize our results in Section \ref{sec:summary}.
The Appendix contains further details about methods for computing the properties of parametric resonance.

\section{Non-Canonical Kinetic Terms}
\label{sec:noncanonKin}

Let us first focus on the implications that non-canonical kinetic terms of the form (\ref{eq:LeffNonCanon}) have on
the motion of the inflaton field oscillating about its minimum.
Inflationary Lagrangians of this form with no potential energy have been proposed as
an alternative to potential-dominated inflation, and are dubbed ``k-inflation" \cite{kinflation}.  
More generally, however, we expect the inflationary sector to have 
both kinetic and potential energy; in this case, the effective Lagrangian ${\mathcal L}_{eff}(X,\phi)$
can be seen as an extension of standard slow-roll inflationary models with non-canonical kinetic terms.
These non-canonical Lagrangians
have a number of interesting properties that make them attractive to inflationary model building and phenomenology.
On the model-building side, the non-canonical kinetic terms significantly modify the dynamics of the inflaton so that
the speed of the inflaton can remain small even when rolling down a steep potential \cite{DBI,Franche:2009gk}.
In many cases, this results in a ``speed limit" for the inflaton, namely a maximum for the speed of the homogeneous
mode of the inflaton.  This is a great advantage to model building, since then it is less necessary to fine-tune the
inflationary potential to have very flat regions supporting slow roll inflation.  
Inflation with non-canonical kinetic terms can also lead to interesting signatures in the CMB,
such as observable non-gaussianities \cite{kinflationPert,DBISky,NonGauss}.

We will choose to work with inflationary Lagrangians of the ``separable" form (discussed in more detail
in \cite{Franche:2009gk,Franche:2010yj}):
\be
{\mathcal L}_{eff}(X,\phi) = \sum_{n\geq 0} c_n \frac{X^{n+1}}{\Lambda^{4n}} - V(\phi) =p(X) - V(\phi)\, ,
\label{eq:separable}
\ee
where we have written $p(X)$ as a power series expansion in $X/\Lambda^4$, with $\Lambda$ some UV energy scale.
Since $\Lambda$ is typically the mass scale of some heavy sector we have integrated out, 
we require it to be larger than the mass of the inflaton $\Lambda \gg m_\phi$ in order for such an effective field theory perspective to make sense.
As a further restriction, we will take $c_0 = 1$, so that for small $X/\Lambda^4 \ll 1$, the Lagrangian reduces to the
usual canonical Lagrangian.  In order for the power series expansion to make sense, the series must have some
non-zero radius of convergence $R$ so that the series converges in the domain of convergence
$X/\Lambda^4 \in [0,R)$, with $R \leq 1$ (it is not necessary that the series itself converge at the boundary).
Finally, we will also require that the first and second derivatives of the power series are positive
$\frac{\partial p}{\partial X}, \frac{\partial^2 p}{\partial X^2} > 0$ so as to guarantee that we satisfy
the null energy condition and have subluminal propagation of perturbations 
\cite{Franche:2009gk,Bean,SickTheories,Babichev:2007dw}.

In a FLRW background
\be
ds^2 = -dt^2 + a(t)^2 d\vec{x}^2
\ee
the scale factor $a(t)$ is driven by the energy density of the homogeneous inflaton $\phi(t)$,
\be
\left(\frac{\dot a}{a}\right)^2 = \frac{1}{3M_p^2} \left[ 2X \frac{\partial p}{\partial X} - p(X) + V(\phi)\right]\, ,
\ee
where now $X = \frac{1}{2}\dot \phi^2$.  The equation of motion for $\phi(t)$ in this background becomes,
\be
\ddot \phi + 3 H \dot \phi c_s^2 + \frac{\partial V}{\partial \phi} \frac{c_s^2}{\partial p/\partial X} = 0
\label{eq:inflatonEOM}
\ee
where the sound speed $c_s^2$, defined by
\be
c_s^2 = \left(1+2X \frac{\partial^2 p/\partial X^2}{\partial p/\partial X}\right)^{-1}\, ,
\ee
is so called because it also appears as the effective speed of perturbations of $\phi$ about this
background.
We will ignore the expansion of the Universe for now, dropping the Hubble friction term in (\ref{eq:inflatonEOM}),
and will return to the effects of expansion of the Universe in Section \ref{sec:expansion}.

For simplicity, we will take the potential $V(\phi)$ of the inflaton to only consist of a mass term
 $V(\phi) = \frac{1}{2}m_\phi^2 \phi^2$.  This could be the entire inflaton potential, as in chaotic inflation,
or just the form of the potential near its minimum.
The difference is not particularly important, as we are
primarily concerned with the phase when the homogeneous inflaton is oscillating about its minimum, where the
quadratic form of the potential will be sufficient. For a scalar field with a canonical kinetic term in this potential, 
the equation of motion is that of a simple harmonic oscillator: the inflaton oscillates sinusoidally 
$\phi(t) = \Phi \sin (m_\phi t)$.

The behavior of a non-canonical kinetic term, however, can be qualitatively different.  As with the canonical case,
the potential provides a force that accelerates the inflaton.  However, now the effective force
$\frac{\partial V}{\partial \phi} \frac{c_s^2}{\partial p/\partial X}$ is also a function of the speed of the inflaton.
Recall that the kinetic term $p(X)$ is a series which is only defined within a finite radius of convergence,
$\frac{1}{2} \frac{\dot\phi^2}{\Lambda^4} \leq R$.  As the inflaton speed approaches the radius
of convergence $|\dot \phi| \rightarrow \sqrt{2R} \Lambda^2$ the series $p(X)$ or its derivatives
$\partial p/\partial X, \partial^2 p/\partial X^2$ may converge or diverge, depending on the precise form of the
series chosen.  Notice that if the second derivative of the series $\partial^2 p/\partial X^2$ diverges faster
than the first derivative $\partial p/\partial X$ at the boundary of the domain of convergence, so that
$c_s^2 \rightarrow \frac{1}{2X} \frac{\partial p/\partial X}{\partial^2 p/\partial X^2} \rightarrow 0$,
then the effective force vanishes.
Said another way, as the force from the potential increases the speed of the inflaton, the non-canonical kinetic
dynamics modify the {\it effective} force felt by the inflaton so that
the effective force, and thus the acceleration of the inflaton, vanish as the inflaton approaches the radius of convergence.
Thus, Lagrangians for which $\partial^2 p/\partial X^2$ diverges as $\dot \phi$ approaches the radius of convergence
have a {\it speed limit}, such that $|\dot \phi| \leq \dot \phi_{max} = \sqrt{2R} \Lambda^2$.
Certainly, as the inflaton approaches the speed limit, we are approaching the boundary
of validity for the EFT (\ref{eq:separable}). Thus, it is advantageous to have a symmetry
that protects the form of the Lagrangian against further corrections as this threshold is
approached \cite{deRham:2010eu}.
Further, requiring perturbivity of the inflaton perturbations places a bound on the minimal
sound speed $c_s^2$, thus restricting how close to the boundary of the EFT we can
go \cite{EFTInflation,PerturbativeInflation,Shandera:2008ai}.
Fortunately, there is a window where both the system is perturbative and the inflaton
is close to its speed limit, so that the effects we are interested in below are still present.

\begin{figure}[t]
\centerline{\includegraphics[width=0.45\textwidth]{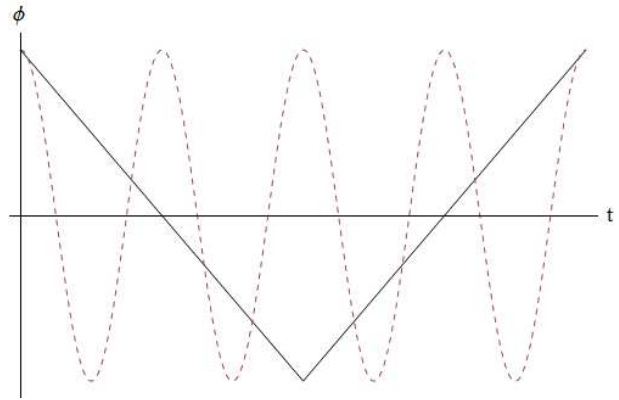}}
\caption{The motion of the homogeneous inflaton $\phi(t)$ about the minimum of the (quadratic) potential
differs depending on whether the kinetic terms
are canonical or non-canonical.  For canonical kinetic terms the motion is sinusoidal.  For non-canonical
kinetic terms the motion approaches a saw-tooth with a much longer period as the inflaton saturates its speed limit.
This form of the inflaton profile is universal for Lagrangians that have a speed limit, except for a very small region
near the turning point which is unimportant for preheating.}
\label{fig:InflatonProfile}
\end{figure}

For potentials that are sufficiently steep, the inflaton quickly attains
its speed limit and stays there until it reaches the other side of the potential, where it decelerates and changes
direction, as shown in Figure \ref{fig:InflatonProfile}.  
This has two important consequences that will play a very important role in preheating:
\begin{enumerate}
\item[(a)] The profile becomes sharper, with a smooth sinusoidal profile turning into a saw-tooth profile.
\item[(b)] The speed limit slows the inflaton down, lengthening the period.
\end{enumerate}
These two effects can be clearly seen in Figure \ref{fig:InflatonProfile}.
As mentioned earlier, the form of the potential about the minimum is in principle unrelated to the form of the potential
during inflation.  This implies that the behavior of the inflaton oscillating about its potential minimum
can be dominated by the non-canonical kinetic terms, independent of whether inflation itself is dominated
by these terms.

When the inflaton saturates the speed limit its profile along one of the legs is approximately linear in time\footnote{
This implies that the acceleration is approximately zero, so higher derivative corrections to the effective
Lagrangian (\ref{eq:separable}) are very small, as discussed in footnote 1.}:
\be
\phi(t) \approx \sqrt{2R}\, \Lambda^2 (t-t_j)\, ,
\label{eq:noncanonprofile}
\ee
where $t_j$ is the time when the inflaton crosses zero.
For small $\Lambda$ this linear approximation of the inflaton is valid up to the turning point $|\phi| \approx \Phi$, so the
period of oscillation is
\be
T_{NCR}= \frac{4\Phi}{\sqrt{2R}\, \Lambda^2}\, ,
\label{eq:NonCanonPeriod}
\ee
with $t\in [-\frac{T_{NCR}}{4},\frac{T_{NCR}}{4}]$.
In order for the non-canonical kinetic terms to be important, this must be much larger than the canonical period
of oscillation $T_{CR} = 2\pi/m_\phi$ (otherwise the inflaton would not reach the speed limit during oscillation), so
\be
T_{NCR}\gg T_{CR} \ \ \Rightarrow \ \ \Lambda \ll \left(\frac{2}{R}\right)^{1/4} \sqrt{\frac{\Phi m_\phi}{\pi}}\, .
\ee
This provides a precise condition for the small $\Lambda$ limit where the system is very non-canonical.
Together with the requirement that the effective field theory description make sense $\Lambda \gg m_\phi$, 
we have the following regime for $\Lambda$ where both the EFT description makes sense and the
speed limit is saturated:
\be
1 \ll \frac{\Lambda}{m_\phi} \ll \left(\frac{2}{R}\right)^{1/4}\sqrt{\frac{\Phi}{\pi m_\phi}}\, .
\label{eq:Lconstraint}
\ee
Clearly, for too small of an amplitude $\Phi$ of the initial oscillation this cannot be satisfied.  Since the expansion
of the Universe and the transfer of energy from the inflationary to the reheating sector cause the amplitude
of oscillation to decrease over time, eventually the inflaton no longer saturates the speed limit
and the inflaton profile behavior returns to the canonical limit.

Perhaps the most well-known non-canonical Lagrangian that leads to a speed-limit for the inflaton is that of
the DBI Lagrangian \cite{DBI} (with constant warp factor, see \cite{Pajer:2008uy}):
\be
p(X)= -\Lambda^4 \left[\sqrt{1-\frac{2X}{\Lambda^4}}-1\right]\, .
\label{DBILagrangian}
\ee
This Lagrangian arises by considering the motion of a space-filling D3-brane in a compact space, with
$\phi$ taking the role of a transverse coordinate of the D3-brane.
In this case the speed limit has a nice
geometrical interpretation - it is just the effective speed of light for motion the extra dimensions
(the effective speed of light is not necessarily unity if the metric on the internal space has non-trivial warping).
We can also represent (\ref{DBILagrangian})  as a power series representation in powers of $X/\Lambda^4$ 
as in (\ref{eq:separable}) with a radius of convergence $R = 1/2$.
But the condition for obtaining the speed limit behavior can be satisfied by a much larger set of Lagrangians,
not just the DBI Lagrangian.  For example, Lagrangians of the form \cite{Franche:2009gk},
\be
p(X) = -\Lambda^4\left[\left((1-\frac{1}{R} \frac{X}{\Lambda^4}\right)^{R} - 1\right],
\label{PLagrangian}
\ee
with $R < 1$ (which includes (\ref{DBILagrangian}) for $R=1/2$), or
\be
p(X) = -\Lambda^4\left[ \log\left(1-\frac{X}{\Lambda^4}\right) -1 \right]
\label{logLagrangian}
\ee
are all of the form (\ref{eq:separable}) (with radii of convergence $R, 1$ respectively) and
all lead to speed-limiting behavior for $\phi$.
Importantly, though, the details of preheating driven by such fields will
be insensitive to these different choices of Lagrangians - as long as there is a speed limit, the profile
$\phi(t)$ will be that of the solid line in Figure \ref{fig:InflatonProfile}.

\section{Preheating with Non-Canonical Inflation}
\label{sec:ncpreheating}

\subsection{Floquet Theory of Resonance}
\label{sec:FloquetResonance}

In the previous section we described the effects of non-canonical kinetic terms on the profile $\phi(t)$ of
an inflaton oscillating about the minimum of its potential, ignoring the coupling
of the inflaton to the reheaton field.  Now, we will consider the impact on preheating of the coupling between the inflaton and reheaton
sectors as in (\ref{eq:twosector}), with the inflationary Lagrangian given by (\ref{eq:separable}),
\be
{\mathcal L}_{pre} = p(X)- \frac{1}{2} m_\phi^2 \phi^2+ \frac{1}{2} (\partial \chi)^2 -\frac{1}{2}m_\chi^2 \chi^2
- \frac{1}{2} g^2 \phi^2 \chi^2 \, ,
\ee
where we will assume that $p(X)$ gives rise to a speed limit.

While the inflaton is dominated by its spatially homogeneous mode $\phi = \phi(t)$, which is oscillating with period $T$, 
the reheaton is assumed to have
a vanishing background vacuum expectation value (VEV) $\langle \chi \rangle = 0$. 
Thus, we will consider fluctuations of the reheaton about the vacuum, which may be decomposed into a set of Fourier modes
\be
\chi = \delta \hat \chi(\vec{x},t) = \int \frac{d^3 k}{(2\pi)^3} \left(e^{i\vec{k}\cdot \vec{x}} \delta \chi_k(t) + e^{-i\vec{k}\cdot \vec{x}} \delta \chi_k^*(t)\right).\ \
\ee
Ignoring for now the expansion of the Universe, the equation of motion for the reheaton fluctuations is
\be
\delta \ddot \chi_k(t) + \left(K + g^2 \phi(t)^2\right)\delta \chi_k(t) = 0
\label{eq:reheatoneom}
\ee
where $K \equiv k^2 + m_\chi^2$.  
This is easily recast into a driven harmonic oscillator known as Hill's equation \cite{HillBook}, after a redefinition
to a dimensionless time coordinate $\tau = 4 \pi (t-t_0)/T$ (a prime denotes a derivative
with respect to $\tau$):
\begin{equation}
\delta \chi_k(\tau)'' + \left[A_k + qF(\tau)\right] \delta \chi_k(\tau) = 0,
\label{hilleq}
\end{equation}
where $F(\tau)$ is a $2\pi$-periodic function symmetric about $\tau = 0$, satisfying 
$\int_{-\pi}^\pi F(\tau) d\tau = 0$.
Floquet's theorem (see \cite{HillBook}) states that solutions to (\ref{hilleq}) have the form
\be
\delta \chi_k(\tau) = e^{\til \mu_k \tau} g(\tau) + e^{-\til \mu_k \tau} g_2(\tau)
\ee
where $g(\tau),g_2(\tau)$ are oscillating solutions, and the Floquet growth exponent $\til \mu_k$  depends on $A_k$ and
$q$ and is complex in general.  For certain ranges of $A_k$ and $q$, known as resonance bands, the real part
of the Floquet growth exponent is non-zero\footnote{Without loss of generality, we will take the real part of the Floquet
exponent to be positive.} leading to exponentially growing solutions:
\be
\delta \chi_k(\tau) \sim e^{\til \mu_k \tau}\, .
\label{floquetSolution}
\ee

In order to compare models of preheating with different oscillating profiles $\phi(t)$ it is more convenient to
parameterize the growth exponent in terms of {\it physical time} $t$:
\be
\delta \chi_k(t) = e^{\mu_k t}
\ee
where $\mu_k = \til \mu_k \frac{2\pi}{T}$, with $T$ the period of oscillation of the inflaton.
The relation between the two growth exponents is that $\til \mu_k$ represents the growth {\it per oscillation}
of the inflaton, while $\mu_k$ represents that growth {\it per unit of physical time}, which takes into account
effects on the overall growth due to changes in the period.
The physical growth exponent $\mu_k$ is the appropriate quantity to use to evaluate the rate of growth compared
to the rate of expansion of the Universe.
Particle production is tracked by the number density $n_k$ of particles, constructed as the energy
of a mode divided by the effective mass $m_{eff,k}^2 = k^2 + m_\chi^2 + g^2 \phi(t)^2$, and
also scales with the Floquet exponent as
\be
n_k = \frac{m_{eff,k}}{2} \left(\frac{|\delta \dot \chi_k(t)|^2}{m_{eff,k}^2} + |\delta \chi_k(t)|^2\right) \sim e^{2\mu_k t}\, .
\ee

For any given oscillating profile $\phi(t)$ for the inflaton, with corresponding $F(\tau)$ in (\ref{hilleq}), it
is possible to numerically determine the resonance bands; the procedure is outlined in Appendix \ref{sec:determinant}.
We distinguish two opposite regimes:
\begin{enumerate}
\item[1)] \textbf{CR}: Canonical Reheating, where the reheaton is coupled to an inflaton with a canonical kinetic term (or equivalently where $\Lambda^2 >> \dot \phi$ at all times so that the inflaton \textit{effectively} behaves canonically), $p(X) \approx X$.
\item[2)] \textbf{NCR}: Non-Canonical Reheating, where the reheaton is coupled to an inflaton with a non-canonical kinetic term $p(X)$, such that the inflaton speed approaches the speed limit $\dot \phi \simeq  \sqrt{2R}\, \Lambda^2$
as it oscillates about its minimum.
\end{enumerate}

\begin{figure*}[t]
 \centerline{\includegraphics[width=0.5\textwidth]{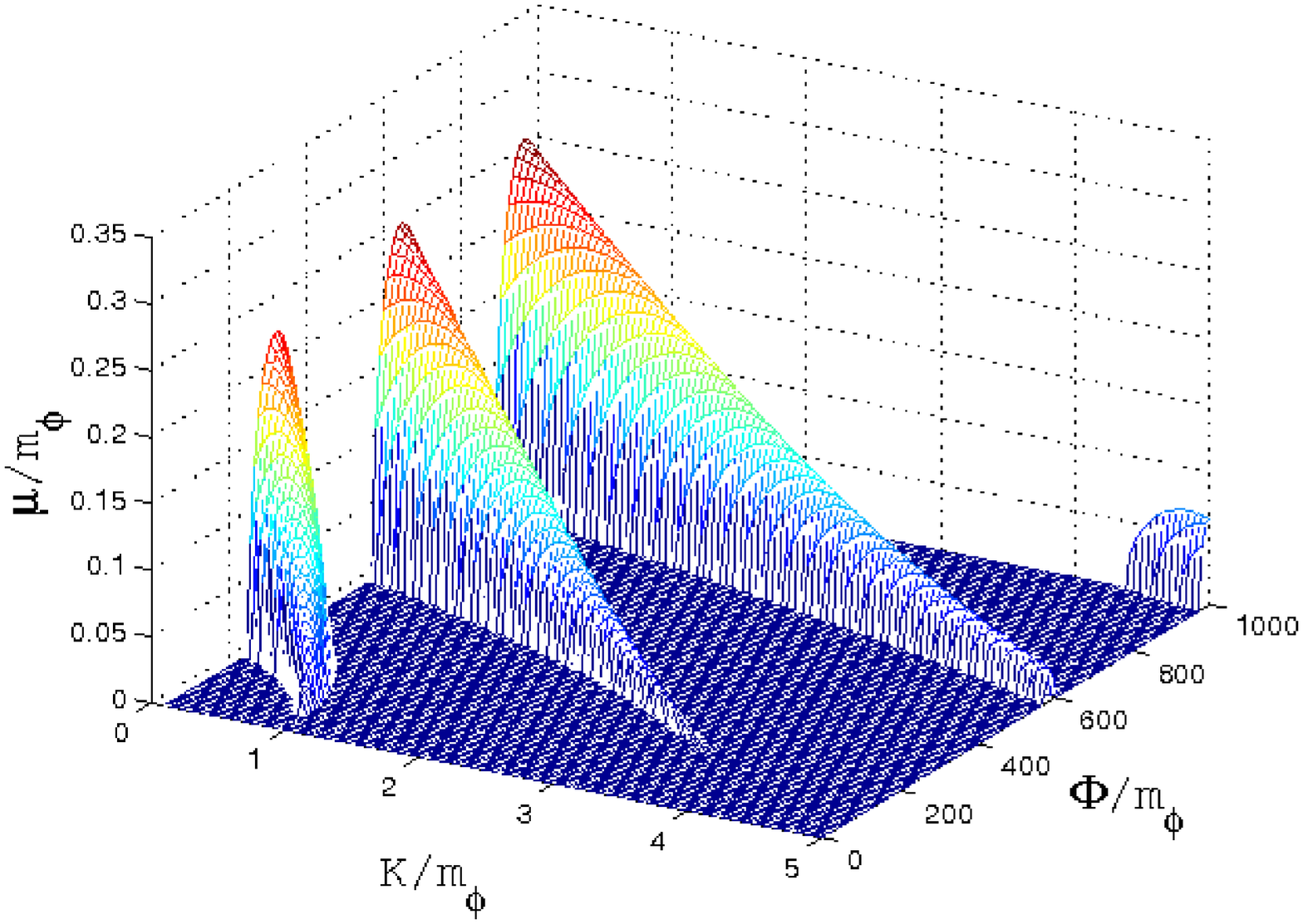} \includegraphics[width=0.5\textwidth]{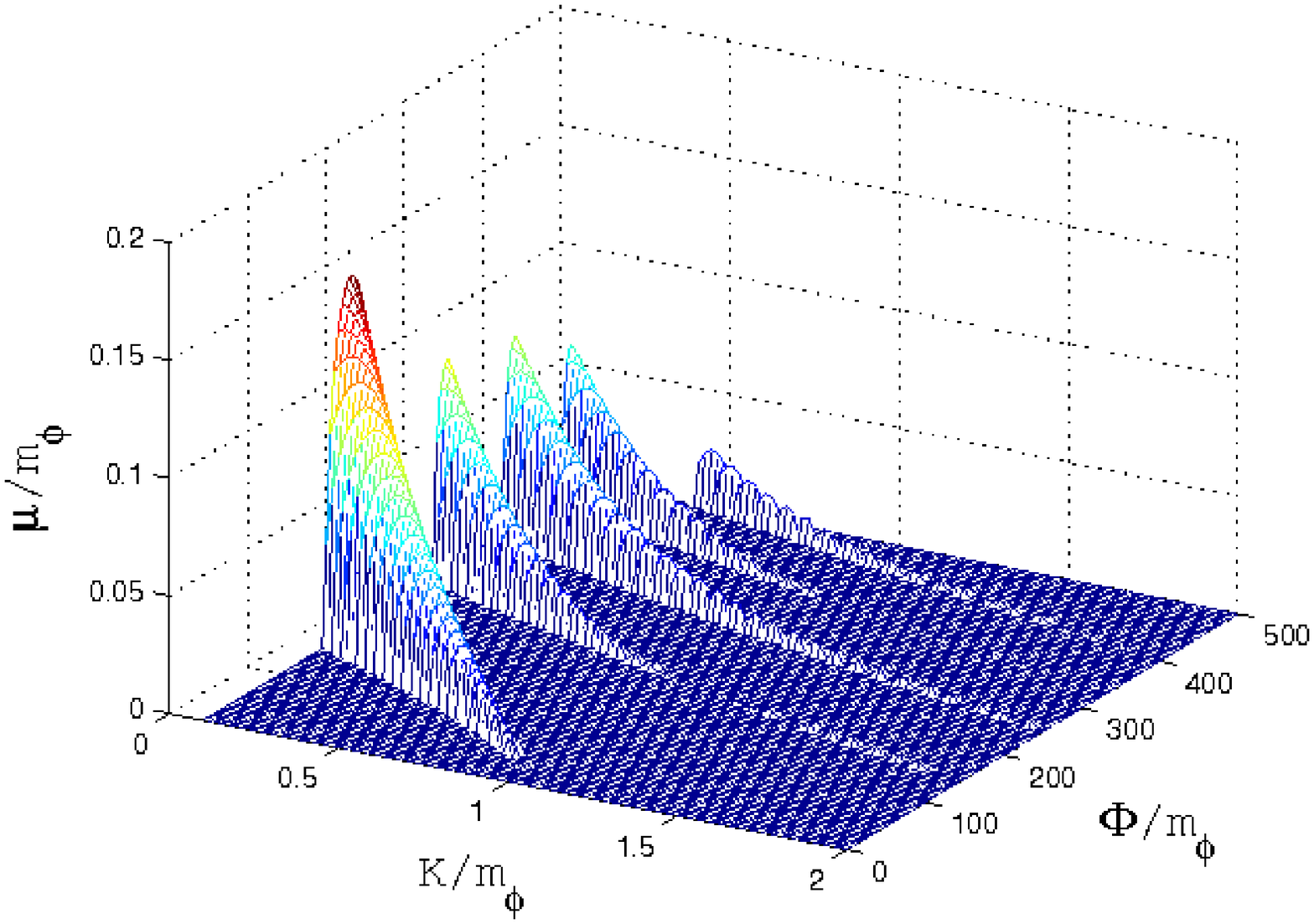}}
 \caption{Resonance bands for reheaton perturbations as a function of the scale $K$ and initial amplitude
 of the oscillating inflaton $\Phi$ with $g=0.005$ in the case when the reheaton is coupled to an 
 inflaton with canonical (CR scenario, left) and
 non-canonical (NCR scenario, right) kinetic terms.}
 \label{fig:resonanceBands}
\end{figure*}

First, let us write the equation of motion for the reheaton fluctuations $\delta \chi_k(t)$ in the form
(\ref{hilleq}) for the CR case.  As discussed in the previous section,
an inflaton oscillating in a quadratic potential with a canonical kinetic term has a sinusoidal profile 
$\phi(t) = \Phi \sin (m_\phi t)$.
The reheaton equation of motion can then be recast into the form of a Hill equation 
(\ref{hilleq}) with the identifications:\be
&&  A_k = \frac{2K + g^2 \Phi^2}{8m_\phi^2},\ \ \  q_{CR} = \frac{g^2\Phi^2}{8 m_\phi^2}, \\
&& \tau = 2 m_\phi t, \ \ \  F(\tau) = \cos \tau
\ee
Hill's equation in this form is more commonly known as the Mathieu equation, and the resonance bands, plotted
in $(K,\Phi)$ space in Figure \ref{fig:resonanceBands}, take the familiar form from previous
studies of preheating \cite{preheating,preheatingLong}.

In the opposite limit (NCR), when the inflaton has a non-canonical kinetic term with a speed limit 
$|\dot \phi|_{max} = \sqrt{2R}\ \Lambda^2$, the profile $\phi(t)$ becomes a  saw-tooth, 
as in Figure \ref{fig:InflatonProfile}.
Along one of the ``legs" of this profile, the inflaton is linear in time as in (\ref{eq:noncanonprofile}), 
so that the reheaton equation of motion becomes:
\be
\delta \ddot \chi_k(t) + \left(K + 2 R g^2 \Lambda^4 t^2\right)\delta \chi_k(t) = 0\, .
\label{eq:chieomnoncanon}
\ee
This can also be rewritten in the form of a Hill's equation (\ref{hilleq}) with the identifications:
\be
&& A_k = \frac{K\Phi^2}{2 R \pi^2  \Lambda^4} + \frac{1}{3}\frac{g^2\Phi^4}{2 R \pi^2 \Lambda^4}, \ \ \ 
  q_{NCR} = \frac{g^2\Phi^4}{2R \pi^2 \Lambda^4}, \nonumber \\
&& \tau = \frac{\sqrt{2 R}\, \Lambda^2 \pi}{\Phi}t,\ \ \  F(\tau) = \frac{\tau^2}{\pi^2} - \frac{1}{3}.
\label{eq:noncanonhillparam}
\ee
The resonance bands for this form of Hill's equation are shown in Figure \ref{fig:resonanceBands}.
The two regimes can easily be connected numerically.

There are two main regimes of interest of (\ref{hilleq}) for the resonance bands: narrow resonance $q \ll 1$ and broad resonance
$q \gg 1$.  
For $g^2 \Phi^2/m_\phi^2 \ll 1$, the CR scenario is in the narrow
resonance regime.  When non-canonical kinetic terms are important, however, the parameter $q_{NCR}$ in the 
corresponding Hill's equation (\ref{eq:noncanonhillparam}) is enhanced relative to the canonical case
\be
q_{NCR} = \frac{g^2}{2R\pi^2} \left(\frac{\Phi}{m_\phi}\right)^2 \frac{m_\phi^2 \Phi^2}{\Lambda^4}\, 
    = q_{CR}\frac{4}{R \pi^2} \frac{m_\phi^2 \Phi^2}{\Lambda^4}.
\ee
The enhancement factor $m_\phi^2 \Phi^2/\Lambda^4 \gg 1$ is large in order for the inflaton to
saturate the speed limit, so unless $q_{CR}$ is correspondingly small, 
when the CR scenario is in the narrow resonance regime the NCR scenario is in broad resonance.
This has important physical implications, since then not only is the growth per period $\til \mu_k$ larger for non-canonical
kinetic terms, but there is also growth over a much larger range of scales.
We will see in the next section, however, that this enhancement is overwhelmed by suppression of particle production due to the lengthening
of the period.

The resonance due to (\ref{eq:reheatoneom}) depends on the physical wavenumber of the fluctuation $k$;
most particle production occurs when the effective mass $K + g^2 \phi^2(t)$ vanishes.  
This implies that the resonance is most
efficient at large scales $K \sim 0$, as can be seen in Figure \ref{fig:resonanceBands}.  
In practice, however, we cannot work on scales larger than the Hubble radius
while neglecting metric perturbations, so $k > H$.  In addition, a non-zero bare mass $m_\chi$ for the reheaton also
keeps the effective mass from vanishing.  But as long as we work on sufficiently large scales (and with a sufficiently
large initial amplitude) so that $K \ll g^2 \Phi^2$, the maximum of resonance at $K = 0$ will be a good approximation
for the maximum resonance at large scales.

\begin{figure}[t]
 \centerline{\includegraphics[width=0.5\textwidth]{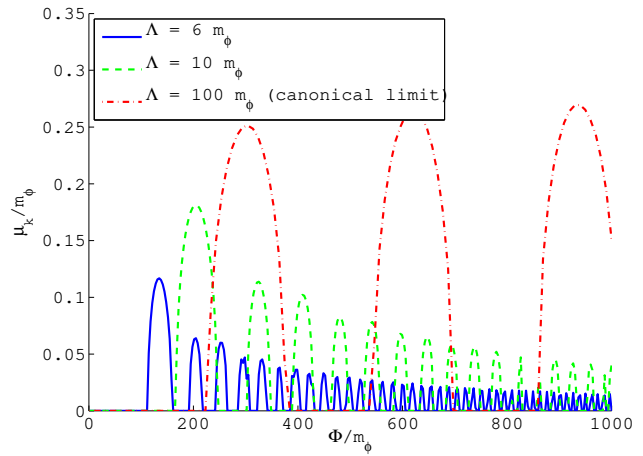}}
 \caption{Floquet growth exponent $\mu_0$ per unit of physical time for $K = 0$ for three different values of $\Lambda$ in DBI inflaton-driven 
 preheating. Lowering the speed limit $\Lambda^2$ of the inflaton greatly reduces the strength of parametric resonance
because the period of oscillation (\ref{eq:NonCanonPeriod}) increases as $1/\Lambda^2$.  The additional suppression
of the resonance due to the dependence of the period on the amplitude of inflaton oscillations $\Phi$ is also evident.}
 \label{fig:muVsPhi}
\end{figure}

Figure \ref{fig:muVsPhi} displays the growth exponent $\mu_k$
as a function of the initial amplitude $\Phi$ of the inflaton for large scales (namely $K=0$).
Several features are evident: for decreasing $\Lambda$, more resonance bands become accessible, because the system
enters the broad resonance regime; the maxiumum size of the growth exponent in the first resonance band 
decreases as $\Lambda$ decreases, reflecting the lengthening of the period; and the heights of the resonance
bands for small $\Lambda$ decrease with increasing $\Phi$, also due to the lengthening of the period for large
initial amplitudes (note that the period for canonical kinetic terms is independent of the amplitude).

\begin{figure*}[t]
\centerline{\includegraphics[width=0.5\textwidth]{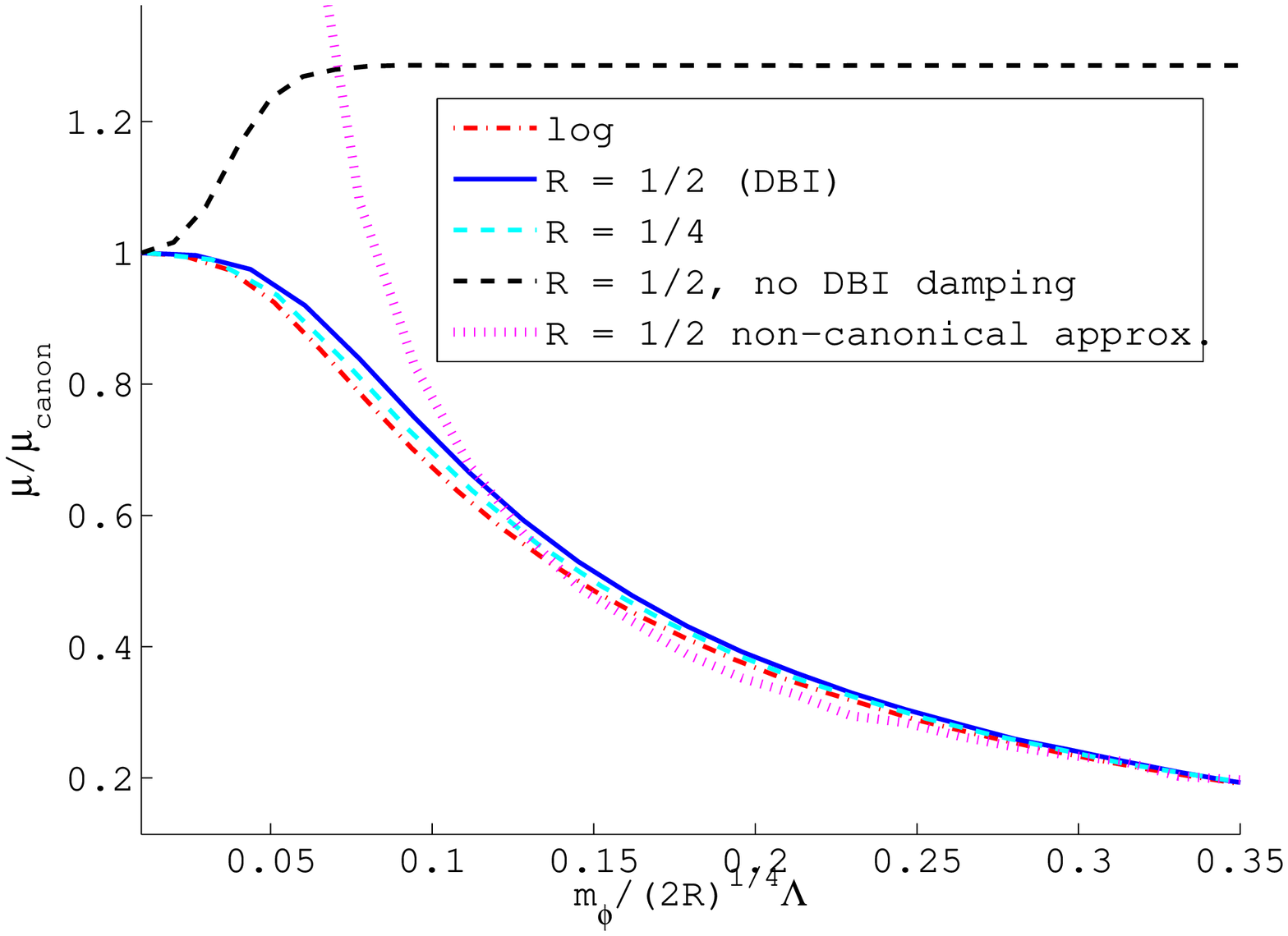}\includegraphics[width=0.5\textwidth]{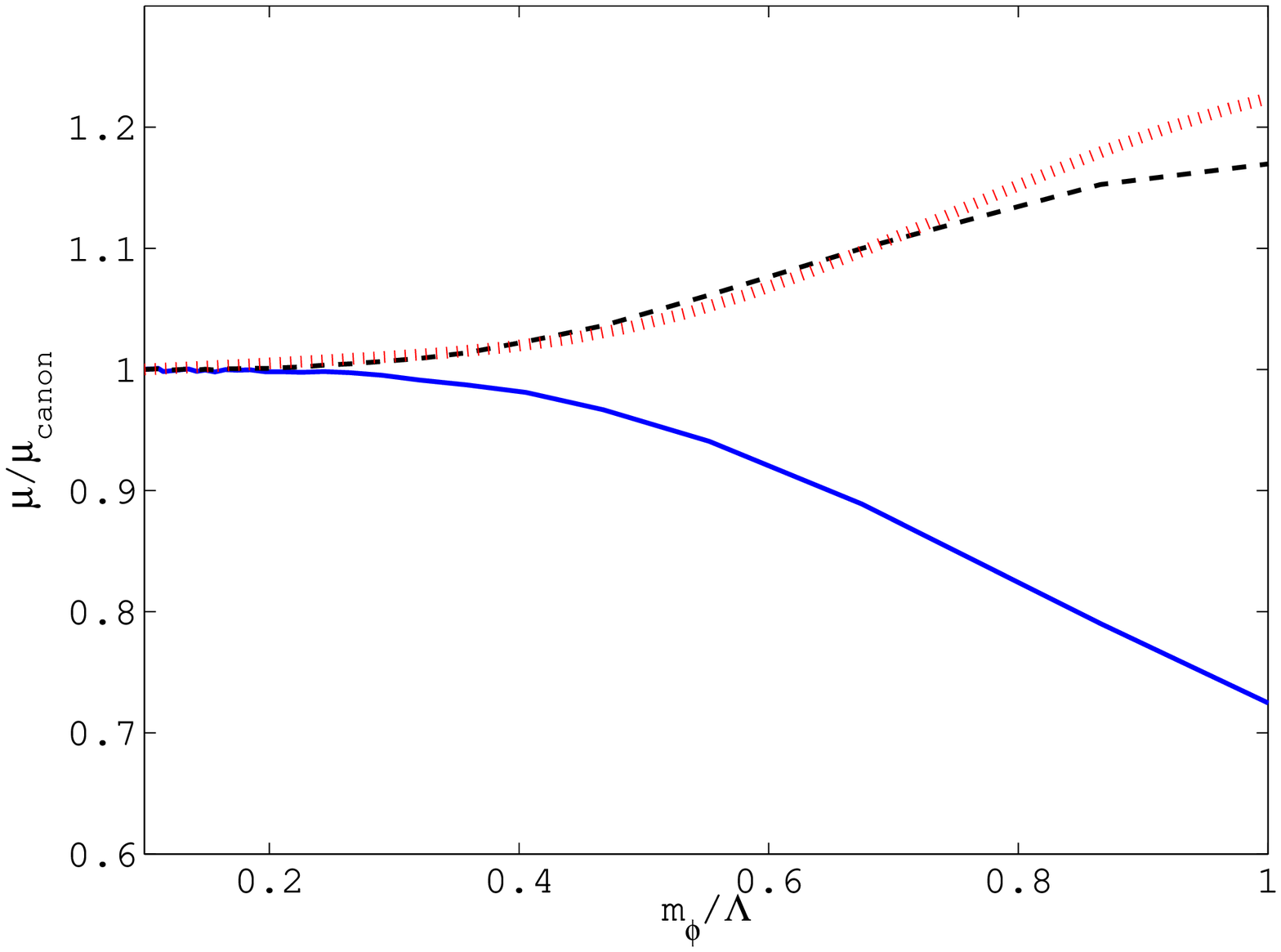}}
\caption{Floquet exponent $\mu$ of the first resonance band for $K = 0$ 
for $g = 0.005$ (left) and $g = 0.5$ (right). $\Lambda >> m_\phi$ corresponds to the canonical case CR. 
Parameter $R$ in the left-hand panel refers to (\ref{PLagrangian}), 
and $\log$ refers to (\ref{logLagrangian}).  
In both panels the black dashed line is the 
growth of particle production per period, whereas the blue solid line is the true growth of particle production per 
physical unit of time. 
Additionally, in the left panel the magenta dotted line is the analytical result (\ref{eq:periodEffect}), showing
good agreement with the numerical results in the regime where the speed limit is important.
In the right panel we have also included the approximate result from \cite{DBIPreheat2} 
(dotted red line, (\ref{muApprox})),
which did not include the suppression from the speed limit. As discussed in Section \ref{sec:noncanonKin}, 
the effective theory breaks down for $m_\phi > \Lambda$ and results should therefore not be trusted beyond this.}
\label{periodEffect}
\end{figure*}

This effect is not limited to the DBI case: Figure \ref{periodEffect} compares the Floquet exponent for several Lagrangians of the forms (\ref{PLagrangian}) and (\ref{logLagrangian}). While the behaviors differ slightly as one moves away from the canonical case, they converge again in the regime where inflaton oscillations saturate the speed limit. The black dotted line in Figure \ref{periodEffect} shows the Floquet exponent when the period lengthening is not taken into account: as the inflaton trajectory approaches the saw-tooth shape, energy is injected into the reheaton field over a longer amount of time, allowing for more particle production. This saturates when the slope $\dot \phi$ approaches the constant speed limit for most of the period. This enhancement is clearly subdominant, however, when compared with the suppression of particle production from the elongation of the period itself.

We close this section by comparing our results with previous explorations in the literature.  Ref.~\cite{DBIPreheat2} 
found an expression for small $\dot \phi/\Lambda^2$ for the growth exponent $\mu_k$ in the case of a DBI inflaton by perturbing the canonical 
equation of motion. Up to order $\dot\phi /\Lambda^2$ and for $K = 0$,  this 
corresponds to:
\begin{equation}
 \mu_k \simeq \sqrt{\left(\frac{\theta_2}{2} \right)^2 - \left(\theta_0^{1/2} - 1\right)^2}
\label{muApprox}
\end{equation}
with
\begin{equation}
 \theta_0 \equiv \frac{g^2 \Phi^2}{2 m^2}\left(1+\frac{9\Phi^2m_\phi^2}{32\Lambda^4} \right), \ \ \ \ \ \ \ \theta_2 \equiv \frac{g^2 \Phi^2}{4 m^2}\left( 1+\frac{3\Phi^2m_\phi^2}{8\Lambda^4} \right).
\end{equation}
This result is illustrated in the right panel of Fig. \ref{periodEffect}. For $m_\phi/\Lambda < 1$ it provides a good approximation of the gain in particle production one would expect from DBI inflation were it not for the lengthening of the period, which the authors of \cite{DBIPreheat2} correctly identified as ``DBI friction'' but did not quantify.
We are interested in the opposite limit, however, where the non-canonical kinetic terms are more than just a small
perturbation.

\subsection{Non-Canonical Limit}

In the non-canonical limit, the equation of motion for the reheaton perturbations (\ref{eq:chieomnoncanon})
can be written in a simple and suggestive form by making a redefinition to a different 
dimensionless time coordinate than we considered earlier:
\be
\tau = \left(g \sqrt{2R} \Lambda^2\right)^{1/2}t = \left(2Rg^2\right)^{1/4}\Lambda\, t\, .
\ee
The reheaton equation of motion then becomes
\be
\frac{d^2 \chi_k(\tau)}{d\tau^2} + \left[\kappa^2 + \tau^2\right] \chi_k(\tau) = 0\, ,
\ee
where $\kappa^2 \equiv \frac{k^2 + m_\chi^2}{\sqrt{2R}\, g \Lambda^2}$.  The time range of the
new time variable is $-\frac{\Delta \tau}{2} \leq \tau \leq \frac{\Delta \tau}{2}$ where
$\Delta \tau = T_\Lambda (2g^2 R)^{1/4} \Lambda = (8 g^2/R)^{1/4} (\Phi/\Lambda)$.

The benefit of making this redefinition is that the problem of broad resonance
can be mapped to the problem of scattering of a particle with energy $\kappa^2$
off of an inverted parabolic potential with an effective Schr\"odinger equation \cite{preheating,preheatingLong}:
\be
\frac{d^2 \chi_k(\tau)}{d\tau^2} + (\kappa^2 - V_{eff}(\tau)) \chi_k(\tau) = 0\, .
\ee
This scattering problem can be solved using standard techniques (see Appendix \ref{sec:parabolic}), leading
to the result:
\be
\til \mu_k = \frac{1}{2\pi} \ln\left[\frac{1+|R_k|^2|}{|D_k|^2} + 2 \frac{|R_k|}{|D_k|^2} \cos \theta_{tot}\right]
\ee
in terms of the reflection and transmission amplitudes $R_k, D_k$ 
and the phase of the reheaton mode $\theta_{tot}$.
For an inverted parabolic potential, the reflection and transmission coefficients are known,
so the average growth index simplifies to be \cite{preheatingLong}:
\be
\til \mu_k = \frac{1}{2\pi} \ln \left[1+2 e^{-\pi \kappa^2}-2\sin \theta_{tot} e^{-\frac{\pi}{2}\kappa^2} \sqrt{1+e^{-\pi \kappa^2}}\right]\, .
\ee
For large scales $\kappa^2\approx 0$ this has a maximum/average value of $\til \mu_k \approx (0.28,0.175)$ \cite{preheatingLong}.

The average growth index $\til \mu_k \approx 0.175$ only gives the typical growth over one period of oscillation for broad resonance, 
and is the
same regardless of whether the inflaton dynamics are canonical or non-canonical.
In particular, it does not depend on the type of non-canonical Lagrangian used - {\it any} non-canonical
Lagrangian that leads to a speed limit for the oscillating inflaton will have the same average growth factor
over one period $\til \mu_k$.
The reason the average
growth index does not depend on whether the system is canonical or non-canonical is because we are working
in the broad resonance limit, where only the slope of the inflaton profile (in dimensionless coordinates) at
the point where the inflaton crosses zero matters.

However, as discussed before, the true comparison of the particle production between the canonical and non-canonical
models should be the growth $\mu_k$ over some fixed {\it physical} time period $\Delta t$
\be
n_k \propto e^{2 \til \mu_k \frac{2\pi}{T}\Delta t} \sim e^{2 \mu_k \Delta t}\, ,
\ee
where $T$ is the period, so that the true growth index is the average growth index over one
period divided by the length of the period, $\mu_k = \til \mu_k 2\pi/T$.  This leads to the main result of this section:
since the period for an oscillating inflaton with non-canonical dynamics is much longer than the period for an
oscillating inflation with canonical dynamics $T_{NCR}\gg T_{CR}$, we have {\it much more growth
in a canonical system} over some fixed physical time
$\Delta t$, $\left(\mu_k\right)_{CR} \gg \left(\mu_k\right)_{NCR}$.
In particular, the ratio between the two should fall off as the ratio of their periods,
\be
\frac{\left(\mu_k\right)_{NCR}}{\left(\mu_k\right)_{CR}} = \sqrt{\frac{R}{2}}\frac{\pi \Lambda^2}{\Phi m_\phi} \ll 1\, ,
\label{eq:periodEffect}
\ee
for decreasing $\Lambda$ at $\kappa = 0$. 
Indeed, this is precisely what is found using numerical techniques, as shown
in Figure \ref{periodEffect}.

Finally, we comment on the possiblity for non-canonical kinetic terms to enhance resonance by turning a putative CR system in narrow resonance into
a NCR system with broad resonance.  A CR system in narrow resonance with $q_{CR} \ll 1$ has $(\mu_k)_{CR} \sim q_{CR} m_\phi$
\cite{preheatingLong}.  As discussed in Section \ref{sec:FloquetResonance}, the same system in NCR has
$q_{NCR} \sim q_{CR} f^2$, with $f = \Phi m_\phi/(\sqrt{2R} \Lambda^2) \gg 1$.
But as just discussed the growth exponent in broad resonance for NCR is a fixed number divided by this enhancement factor,
$(\mu_k)_{NCR} \sim 0.1 m_\phi/f$.  Thus, in order for the NCR system to be in broad resonance $q_{NCR}\, \gsim\, 1$, we have that
$(\mu_k)_{NCR}\, \lsim\, 0.1 m_\phi q_{CR}^{1/2}$, representing an enhancement from CR narrow resonance by a factor of $0.1/q_{CR}^{1/2}$.
While this enhancement can be large if the original CR system is very far in the narrow resonance regime, the growth factor is still
quite suppressed.

\subsection{Expansion of the Universe}
\label{sec:expansion}

Up to now we have ignored the expansion of the Universe.  Including expansion has several effects on
preheating which have been well-studied for CR \cite{preheating,preheatingLong,PreheatExpanding}:
\begin{itemize}
\item[i)] The amplitude of inflaton oscillations decreases due to Hubble damping inversely proportional to time 
$\Phi(t) \sim t^{-1}$.
\item[ii)] The parameters of Hill's equation $(A_k, q)$ become time dependent.  In particular, $q_{CR}\sim t^{-2}$
due to the time dependence of the inflaton amplitude.
\item[iii)] The physical wavenumber $k_{phys} = k/a(t)$ redshifts with the expansion, so each comoving mode
only spends a finite amount of time in a resonance band.
\item[iv)] The resonance is {\it stochastic} \cite{preheatingLong}.
\item[v)] Preheating is ineffective when the rate of particle production drops below the expansion rate, e.g.~when
$\mu_k < H$.  For narrow resonance this condition cannot be satisfied, thus particle production can only occur in the
broad resonance regime.
\end{itemize}
We will now make some simple estimates of how these effects will change for NCR.

\begin{figure}[t]
\centerline{\includegraphics[width=0.5\textwidth]{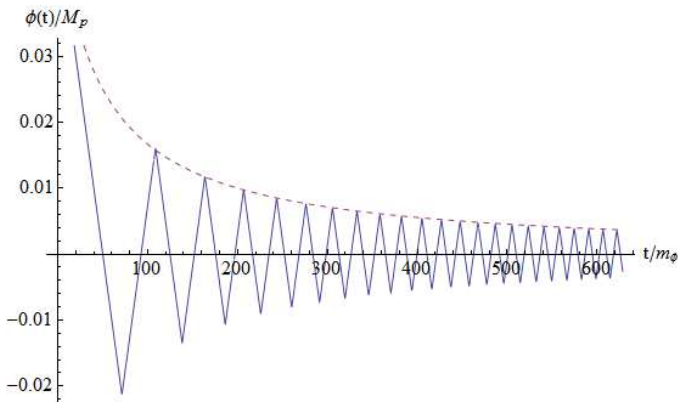}}
\caption{Expansion of the Universe causes the amplitude of the inflaton to decay inversely proportional to time,
as shown here for $\Lambda = 10\, m_\phi$ and $m_\phi = 10^{-5} M_p$ with an initial amplitude of $\Phi_0 = 0.05 M_p$.  The dashed line indicates the $t^{-1}$ behavior of the decaying amplitude.}
\label{fig:Inflaton_Decay}
\end{figure}

First, the Hubble friction term may be removed from the inflaton equation of motion (\ref{eq:inflatonEOM}) by rescaling the inflaton field by powers of the scale factor $\phi(t) = \Phi_0 \til \phi(t)/a(t)^{3/2c_s^2}$, so that
$\til \phi(t)$ is a fixed unit-amplitude oscillating function, and $\Phi(t) = \Phi_0/a(t)^{3/2c_s^2}$ is the time-dependent
amplitude.
When the period of oscillation of the inflaton is much smaller than the timescale for the expansion of the Universe, the energy density can be averaged over several oscillations $\rho_\phi \sim m_\phi^2 \phi^2 \sim a(t)^{-3/c_s^2}$.
The scale factor thus grows as $a(t) \sim t^{2/3 c_s^2}$, so that the amplitude falls is inversely proportional to time
$\Phi(t) \sim \Phi_0/t$ as with a canonical kinetic term.  The decaying inflaton profile in an expanding background is
shown in Figure \ref{fig:Inflaton_Decay}.
It is worth noting that in the far non-canonical limit the period of the inflaton can be larger than the Hubble time due
to the period lengthening effect.  When this occurs, Hubble friction damps a significant amount of the amplitude
in the first oscillation, quickly bringing the system back to the previous case.

For a fixed cutoff scale $\Lambda$, the condition for the oscillating inflaton to saturate the speed limit 
(\ref{eq:Lconstraint}) is violated once the amplitude of oscillation drops below some critical amplitude.
The number of oscillations of the inflaton scales with time $N \sim t/T$, so that the time-dependent amplitude
drops below this critical amplitude after $N \sim m_\phi \Phi_0/\Lambda^2$ oscillations, where $\Phi_0$
is the initial amplitude.  Thus the inflaton no longer saturates the speed limit after
$N \sim m_\phi \Phi_0/\Lambda^2$ oscillations, and behaves canonically after this time.

Another important effect is the dependence of the Hill's equation parameters (\ref{eq:noncanonhillparam}) on
time.  Since $q_{NCR}$ depends on the fourth power of the amplitude, it quickly becomes small at a much faster
rate $q_{NCR} \sim t^{-4}$ than it would in CR, $q_{CR} \sim t^{-2}$.  In particular, after only
$N \sim g^{1/2} \Phi_0/\Lambda$ oscillations, the narrow resonance regime where $q_{NCR}\, \lsim\, 1$ is attained.
As noted above, narrow resonance with CR is inefficient for an expanding background.  Since we argued
in the previous section that particle production for NCR is suppressed compared to CR, narrow resonance
for NCR is also inefficient in an expanding background.  This means that after $N \sim g^{1/2} \Phi_0/\Lambda$ 
oscillations the resonance shuts off due to expansion of the Universe.
The other effects of an expanding Universe listed above for CR (iii,iv) do not appear to be qualitatively different
for NCR, so we will not comment further on them.

In summary, including expansion of the Universe for NCR leads to a decaying amplitude of inflaton oscillations inversely
proportional to time $\Phi(t) \sim t^{-1}$.  After many oscillations the amplitude decreases so much that either:
(a) the inflaton ceases to saturate the speed limit during oscillation, effectively becoming canonical; or (b) the
Hill's equation parameter $q_{NCR}(t)$ becomes so small that particle production cannot compete with
the expansion of the Universe, shutting off the resonance.  Which outcome dominates depends on the relative
magnitudes of $m_\phi/\Lambda$ and $g^{1/2}$; when the former dominates, outcome (a) occurs first, 
and \textit{vice-versa}.

\section{Conclusion}
\label{sec:summary}

Preheating in the post-inflationary Universe is the explosive production of particles far from thermal equilibrium that 
occurs as the inflaton field oscillates about its potential minimum.
We have examined preheating for an inflaton sector that has non-canonical kinetic terms arising as an effective 
theory ${\mathcal L}_{eff}(X,\phi)$, for $X = \frac{1}{2}(\partial \phi)^2$, 
which may arise from the existence of new physics at some energy scale $\Lambda > m_\phi$. 
Effective theories of this type can give rise to a speed limit for the motion of the inflaton $\phi$, 
as in DBI inflation \cite{DBI}.
In addition to having important implications for inflationary model building, the speed limit plays an important role
in modifying the nature of preheating in the post-inflationary Universe.
In particular, as the non-canonical terms become important, non-canonical preheating departs 
significantly from the canonical case via three main effects:
\begin{itemize}
\item[i)]The sinusoidal inflaton profile becomes a saw-tooth, elongating the fraction of the inflaton period in which significant particle production may occur, and moves the system from narrow to broad resonance.
\item[ii)]Effect \textit{i)} is offset by an elongation of the inflaton oscillation period by a factor $f = \sqrt{\frac{2}{R}} \frac{\Phi m_\phi}{\pi \Lambda^2} \gg 1$. This suppresses the amount of $\chi$ particle production \textit{per unit time} by $1/f$.
\item[iii)]Effect \textit{ii)} affects the competition between $\chi$ production and Hubble expansion, making preheating
even less efficient in an expanding Universe.
\end{itemize}
In general, then, preheating when the inflation has non-canonical kinetic terms is {\it less efficient} than with a purely
canonical kinetic term.  This implies that if preheating is to be important at all in the early Universe, then the UV scale
of new physics that couples kinetically to the inflaton must be sufficiently high that the effective non-canonical kinetic 
terms are negligible.

We have only focused on non-canonical kinetic terms for the inflaton sector in this paper.  Certainly, however,
due to the non-linear nature of the parametric resonance, it would be interesting to study how non-canonical
kinetic terms for the reheating sector would affect the physics of preheating.  We leave this to future work.


\section*{Acknowledgments}

We would like to thank Robert Brandenberger and Tomislav Prokopec for useful conversations.
B.U~is supported in part by NSERC, an IPP (Institute of Particle Physics, Canada) Postdoctoral Fellowship, and
by a Lorne Trottier Fellowship at McGill University. ACV was supported by FQRNT (Quebec) and NSERC (Canada). 

\appendix

\section{Computing the Floquet exponent}
\label{sec:determinant}

There exists a straightforward method of computing the Floquet exponent $\mu_k$ as a function of the parameters $A_k$ and $q$, which has been extensively covered in the literature (e.g. \cite{HillBook,Charters:2008es,Lachapelle:2008sy}).

We start with Hill's equation of the form (\ref{hilleq}). The periodic function $F(\tau)$ may be decomposed:
\begin{equation}
 F(\tau) = \sum_{n=-M}^{M} d_n e^{in\tau},
\end{equation}
with $A_k$ defined such that:
\begin{equation}
 \int_{-\pi}^{\pi}F(\tau)d\tau = 0.
\end{equation}
Floquet's theorem (see \cite{HillBook}) states that solutions to (\ref{hilleq}) are of the form,
\begin{equation}
\chi_k(\ta) = e^{\til \mu_k \ta} g(\ta)+e^{-\til \mu_k \ta} g_2(\ta)
\label{eq:Floquetsolns}
\end{equation}
where $g(\ta)$ and $g_2(\ta)$ are periodic functions with period $T$, and $\til \mu_k$, called the {\it Floquet exponent}
or {\it characteristic exponent}, is complex.  Clearly when $\til \mu_k$ has a non-zero real part we have
exponential growth of $\chi_k$ --- this is the parametric resonance effect.  Without loss of generality, we will
take the real part of $\til \mu_k$ to be positive, and so we will drop $g_2(\tau)$ since its coefficient is exponentially
decreasing.

In order to find solutions of the form (\ref{eq:Floquetsolns}), we first Fourier expand $g(\tau)$
\begin{equation}
\chi_k(\tau) 
  = \sum_{n=-\infty}^{\infty} c_n e^{(\til \mu_k + in)\tau}
\end{equation}
and plug this Fourier series back into the equation of motion (\ref{hilleq}) to derive a recursion relation
for the coefficients $c_n$ in terms of $\til \mu_k, A_k, q, d_n$:
\begin{equation}
c_n + \frac{q\sum_{m=-M}^M d_m c_{n-m}}{\left((\til \mu_k+in)^2+A_k\right)} = 0\ \ \ \ \forall n\in (-\infty,\infty)\, .
\label{eq:HillRecursion}
\end{equation}
These recursion relations define a matrix problem
\begin{equation}
B(\til \mu_k,A_k,q,d_m) \begin{pmatrix}c_{-n} \cr \vdots\cr c_{-1} \cr c_0 \cr c_1 \cr \vdots\cr c_n \end{pmatrix} = 0
\label{eq:RecursionMatrix}
\end{equation}
with $n\rightarrow \infty$, where the elements $B_{rs}$ of the (infinite) matrix $B$ are given by:
\begin{equation}
B_{rs} = \begin{cases} 1 & \mbox{if $r=s$} \cr \frac{q d_{r-s}}{(\til \mu_k+ir)^2+A_k} & r\neq s \end{cases}\, .
\end{equation}

The matrix problem (\ref{eq:RecursionMatrix}) implies that the (infinite) matrix $B$ is singular, 
so it must have vanishing determinant
\begin{equation}
\Delta(\til \mu_k,A_k,q,d_m) = \left|B_{rs}\right| = 0\, .
\label{eq:deltaDef}
\end{equation}
The vanishing of the determinant (\ref{eq:deltaDef}) defines an implicit function $\til \mu_k = \til \mu_k(A_k,q,d_m)$.
In practice, we perform a Fourier transform of the inflaton profile written in terms of $F(\tau)$ numerically, and evaluate
the $M\times M$ determinant (\ref{eq:deltaDef}); a matrix size $M \sim 100$ is sufficient for convergence of the
first tens of resonance bands.

Let us briefly comment on the differences in solving for the growth exponent $\til \mu_k$ using this method for
canonical and non-canonical kinetic terms.
Clearly, $F(\tau)$ for non-canonical kinetic terms (\ref{eq:noncanonhillparam}) expanded
in Fourier modes involves many terms, as opposed to the single Fourier mode for canonical kinetic terms.
Thus, one difference between resonance with canonical and non-canonical kinetic terms is the inclusion of more
terms in the Fourier expansion of the profile. 
Naively, then, there are more modes in the driving force with which the $\chi_k$ fields may resonate.   This is the feature arising from the ``sharpening" of the profile.  However, the other important effect of the non-canonical kinetic terms
is the extreme lengthening of the period of oscillation, thus suppressing the resonance in physical time.

\section{Scattering from a Parabolic Potential}
\label{sec:parabolic}

\begin{figure*}[t]
\centerline{\includegraphics[width=.9\textwidth]{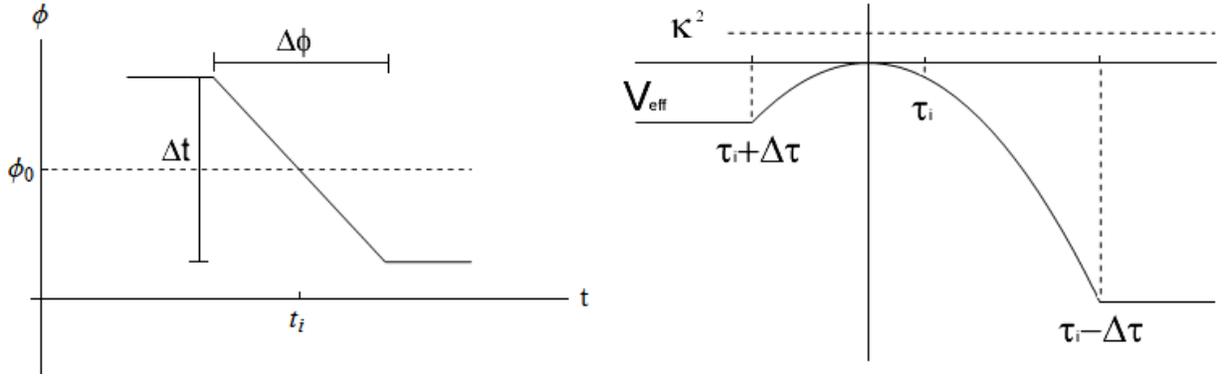}}
\caption{Left: Linearizing $\phi(t) \approx \phi_0 + \lambda (t-t_i)$ about some point $(t_i,\phi_0)$.  
Right: The associated effective scattering problem.}
\label{fig:scatteringsetup}
\end{figure*}

In many cases of interest, the equation for perturbations of the reheaton $\chi_k$:
\ba
\ddot \chi_k(t) + \left[k^2 + m_\chi^2 + g^2 \phi^2\right]\chi_k(t) = 0,
\ea
can be converted into a Schr\"odinger scattering problem.
In this appendix, we will outline the equivalence between these two perspectives; see \cite{preheating}
for more on the correspondence.
We will imagine that $\phi(t)$ has a large slope in some region for some finite amount of time, as
shown in Figure \ref{fig:scatteringsetup}.
Linearizing about this point, we have
\ba
\phi \approx \phi_0 + \lambda (t-t_i)\, .
\ea
This linear approximation is valid roughly for some time interval $2 \Delta t$, i.e. for
$t_i - \Delta t \leq t \leq t_i + \Delta t$. Outside this interval, the field $\chi$ must behave adiabatically:
its potential is slowly-varying and can therefore be approximated by a WKB approach.
The equation of motion around the point of adiabaticity violation becomes:
\ba
\ddot \chi_k(t) + \left[k^2 + m_\chi^2 + g^2 (\phi_0 + \lambda (t-t_i))^2\right]\chi_k(t) = 0\,.
\ea
We can map this problem into a Schr\"odinger-like equation by redefining the time variable
$\tau' \equiv \left(\frac{g}{\lambda}\right)^{1/2}(\phi_0 + \lambda(t-t_i))$, so that the equation
of motion is written:
\ba
\frac{d^2 \chi_k(\tau')}{d\tau'^2} + \left[\kappa^2 + \tau'^2 \right]\chi_k(\tau') = 0.
\label{eq:ScattEq}
\ea
We have defined $\kappa^2 \equiv \frac{k^2 + m_\chi^2}{\lambda g}$.  The time range of the new variable
is
$\tau_i' - \Delta \tau' \leq \tau' \leq \tau_i' + \Delta \tau'$, where $\tau_i' = \phi_0 (g/\lambda)^{1/2}$ and
$\Delta \tau' = (\lambda g)^{1/2} \Delta t$.
The transformed equation (\ref{eq:ScattEq}) resembles a Schr\"odinger equation with the effective potential:
\begin{widetext}
\ba
V_{eff} = \begin{cases}
-(\tau_i'-\Delta\tau')^2 =-V_0 = \mbox{const} & \tau' < \tau_i' - \Delta \tau' \mbox{ (Region I) }\\
 -\tau'^2 & \tau_i' - \Delta \tau' \leq \tau' \leq \tau_i' + \Delta \tau'\mbox{ (Region II) }\\
-(\tau_i'+\Delta \tau')^2 =-V_1 = \mbox{const} & \tau' > \tau_i' + \Delta \tau' \mbox{ (Region III) }
\end{cases}
\ea
\end{widetext}
The setup of this effective scattering problem is shown in Figure \ref{fig:scatteringsetup}.

The solution in each region is therefore known:
\ba
\label{chi1}
\chi_k = \begin{cases} A_1 e^{-ik_1\tau'} + B_1 e^{+ik_1\tau'} & \mbox{(Region I)}\\
A_2 D_\nu\left((1+i)\tau'\right) + B_2 D_{\bar\nu} \left((-1+i)\tau'\right)\hspace{-.1in} & \mbox{(Region II)}\\
A_3 e^{-ik_2 \tau'} +  B_3 e^{+ik_2 \tau'} &\mbox{(Region III)}  
\end{cases}
\ea
where $k_1 \equiv \sqrt{\kappa^2 + V_0^2}$, $k_2 = \sqrt{\kappa^2+V_1^2}$,
$\nu \equiv \frac{1}{2} i (i-\kappa^2)$, and the functions $D_\nu(x)$ are {\it parabolic cylinder functions}.

In the language of scattering matrices, the ingoing/outgoing waves are parameterized as
\ba
\begin{pmatrix}
\alpha_k^{j+1} e^{-i\theta_j^k} \\
\beta_k^{j+1} e^{i\theta_j^k}
\end{pmatrix} =
\begin{pmatrix}
\frac{1}{D_k} & \frac{R_k^*}{D_k^*} \\
\frac{R_k}{D_k} & \frac{1}{D_k^*}
\end{pmatrix}
\begin{pmatrix}
\alpha_k^j e^{-i\theta_k^j} \\
\beta_k^j e^{i\theta_k^j}
\end{pmatrix}
\label{eq:scatM}
\ea
where $D_k,R_k$ are the transmission and reflection coefficients, the $\alpha_k,\beta_k$ are normalized
such that $|\alpha_k|^2-|\beta_k|^2 = 1$ with the ``particle number" defined as $n_k = |\beta_k|^2$,
and $\theta_k^j \equiv \int_0^T \sqrt{k^2 + m_\chi^2 + g^2 \phi^2} dt$ is the accumulated WKB phase
up to the time of scattering. Apart from an overall normalization, it should be clear that the $\alpha^j$'s correspond to the $A_j$'s and the $\beta^j$'s to the $B_j$'s above. 

From (\ref{eq:scatM}) we can write
\ba
\beta_k^{j+1} = \alpha_k^j e^{-2i\theta_k^j} \frac{R_k}{D_k} + \beta_k^j \frac{1}{D_k}
\ea
which we can use to find the final number of particles $n_k^{j+1}$ in terms of the original number of particles $n_k^j$
(together with the normalization condition on the $\alpha$):
\ba
n_k^{j+1} &=& \left|\frac{R_k}{D_k}\right|^2 + \frac{1+|R_k|^2}{|D_k|^2} n_k^j
\nonumber \\
&&   + 2 \frac{|R_k|}{|D_k|^2} \cos \theta_{tot} \sqrt{(1+n_k^j)n_k^j} 
\ea
where $\theta_{tot} \equiv \mbox{arg}\, \alpha_k^j - \mbox{arg}\, \beta_k^j - \mbox{arg}\,R_k - 2\theta_k^j$
is the total accumulated phase.
In the large $n_k^j$ limit, we can drop the first term and simplify the last term, so that we have
\ba
n_k^{j+1} = \left[\frac{1+|R_k|^2}{|D_k|^2} +2 \frac{|R_k|}{|D_k|^2} \cos \theta_{tot} \right]n_k^j\, .
\ea
The average growth exponent over one period is defined as
\ba
n_k^{j+1} = e^{2\til \mu_k \frac{\pi}{T} \Delta t} n_k^j = e^{2\pi \til \mu_k} n_k^j
\ea
for $\Delta t = T$ over one period.  Thus, the growth exponent is,
\ba
\label{muk}
\til \mu_k = \frac{1}{2\pi} \log\left[\frac{1+|R_k|^2}{|D_k|^2} +2 \frac{|R_k|}{|D_k|^2} \cos \theta_{tot} \right].
\ea
Averaging over many possible initial phases, we can consider $\theta_{tot}$ as
a random variable; thus, we will take $\cos \theta_{tot} \sim 0$ on average.

In order to compute the scattering amplitudes $R_k, D_k$, one needs to solve the continuity equations
across the boundaries of Regions I, II and III for the wavefunctions (\ref{chi1}).  In general, it is not
possible to solve these equations analytically.  However, for special cases the equations can
in fact be solved.

The first limit we will take is $|\tau_i'| \gg 1$; in the original setup, this corresponds to taking 
$\phi_0 \sqrt{g}/\sqrt{\lambda}$ to
be large, e.g.~the region of linearization takes place very far from $\phi = 0$.
In this limit the effective potential becomes essentially a step
function:
\ba
V_{eff} = \begin{cases}
-V_0 & \tau' < \tau_i'  \mbox{ (Region I) }\\
-V_1 & \tau' > \tau_i' \mbox{ (Region II) }
\end{cases}
\ea
where $|V_0-V_1| = \sqrt{g/\lambda} |\Delta \phi|$, with $\Delta \phi$ the change in the inflaton over the
time period $\Delta t$, and $|V_0|\sim |V_1| \gg 1$.
In order for the step function approximation to be a valid approximation to the inverted quadratic effective
potential, we need that the change over the step function be small, e.g.~$\frac{|V_0-V_1|}{V_1} \ll 1$.
The transmission and reflection coefficients can easily be computed
\ba
D_k^2&=&\frac{4 (\kappa^2+V_0)(\kappa^2+V_1)}{(2\kappa^2 + V_0+V_1)^2};\nonumber \\
R_k^2&=&\frac{(V_0-V_1)^2}{(2\kappa^2+V_0+V_1)^2};  \label{DkRk}
\ea
so that the growth exponent becomes
\ba
\til \mu_k = \frac{1}{2\pi} \ln\left[\frac{1+(V_0-V_1)^2}{4 \sqrt{(\kappa^2+V_0)(\kappa^2+V_1)}}\right]\,.
\ea
With the requirements above that the size of the step be small, we have $D_k \approx 1$ and $R_k \approx 0$,
so that $\til \mu_k \sim \ln (1) \sim 0$.  This is expected - if the inflation has a very large offset, the effective mass
of the reheaton $m_{eff}^2 = k^2 + m_\chi^2 + g^2 \phi_0^2$ is large, so there will be very little particle production.

Alternatively, in the limit $\tau_i' = 0$, the scattering potential becomes symmetric about $\tau = 0$.
For $\Delta \tau \gg 1$ the scattering amplitudes are known \cite{preheating}:
\be
D_k &=& \frac{e^{i\varphi_k}}{\sqrt{1+e^{-\pi \kappa^2}}}; \label{DkeqKLS}\\
R_k &=& - \frac{i e^{i\varphi_k}}{\sqrt{1+e^{\pi \kappa^2}}};\label{RkKLS}
\ee
where the angle $\varphi_k$ is
\be
\varphi_k = \mbox{arg}\Gamma\left(\frac{1+i\kappa^2}{2}\right) + \frac{\kappa^2}{2}\left(1+\ln \frac{2}{\kappa^2}\right)\,.
\ee
The implications of scattering in this case is considered in more detail in the main text.

\bibliography{GeneralizedReheating}

\end{document}